\documentclass[11pt]{article}
\usepackage{graphicx}
\usepackage{amssymb}

\input{tcilatex}

\begin{document}

\title{Gel Electrophoresis of DNA Knots in Weak and Strong Electric Fields.}
\author{C. Weber$^{1}$, A.Stasiak$^{2}$, M. Fleurant$^{3}$, P. De Los Rios$%
^{3}$ \\ and G. Dietler$^{4,\ast }$ \\
$^{1}$IRRMA, EPFL, Switzerland \\
$^{2}$LAU, UNIL, Switzerland \\
$^{3}$Institut de Physique Th\'{e}orique, EPFL, Switzerland\\
$^{4}$Laboratoire de Physique de la Mati\`{e}re Vivante-IPMC,\\
Facult\'e des Sciences de Base,\\
Ecole Polytechnique F\'ed\'erale de Lausanne,\\
CH-1015 Lausanne, Switzerland.\\
}
\date{\today}
\maketitle

\begin{abstract}
Gel electrophoresis allows to separate knotted DNA (nicked circular) of
equal length according to the knot type. At low electric fields, complex
knots being more compact, drift faster than simpler knots. Recent
experiments have shown that the drift velocity dependence on the knot type
is inverted when changing from low to high electric fields. We present a
computer simulation on a lattice of a closed, knotted, charged DNA chain
drifting in an external electric field in a topologically restricted medium.
Using a simple Monte Carlo algorithm, the dependence of the electrophoretic
migration of the DNA molecules on the type of knot and on the electric field
intensity was investigated. The results are in qualitative agreement with
electrophoretic experiments done under conditions of low and high electric
fields: especially the inversion of the behavior from low to high electric
field could be reproduced. The knot topology imposes on the problem the
constrain of self-avoidance, which is the final cause of the observed
behavior in strong electric field.
\end{abstract}

\eject

\eject

Gel electrophoresis of linear and circular DNA and its dynamics has been
since long time a topic on which numerical simulations and analytical models
have been applied \cite%
{Noolandi87,Deutsch88,Viovy88,Doi88,Noolandi89,Viovy92,Aalberts95,Joanny97}.

Most experimental and theoretical studies of the electrophoresis process
deal with linear or circular DNA \cite{Schwartz89,Smith89,Sturm89,Chu89}.
But DNA comes also in knotted form. Various classes of enzymes
(topoisomerases and site-specific recombination enzymes) produce different
types of knots or catenanes by acting on circular DNA molecules \cite%
{cozz,spengler}. The analysis of these knots gives some information about
the mechanisms by which these enzymes are involved in the proper functioning
of chromosomes (see for example \cite{sikorav}) and about DNA packing \cite%
{roca02}. Being able to study which knots are produced by a given enzyme in
prescribed conditions implies being able to perform some sort of "knot
spectroscopy", which can be done for example by electron microscopy, where
knots are observed one by one. Yet, if large numbers of knots need to be
classified, then some high throughput technique is needed. Such a technique
is gel electrophoresis. Indeed, experimental work has shown a linear
relationship between the distance of electrophoretic migration on agarose
gel of different types of DNA knots (all with the same number of base pairs)
and the average crossing number of the ideal geometrical representations of
the corresponding knots (closely related to the complexity of the knot)\cite%
{stas96}. As a consequence, the type of a knot can be simply identified by
measuring its position on the gel, without the need of electron microscopy
experiments as required until recently.

At low electric field the usual observation is that the more complex the
knot is, the higher is its mobility. A simple intuitive explanation for this
behavior is that the compactness of a knot increases with its complexity
(for a constant string length) and the friction coefficient $\gamma = 6 \pi
\eta_0 R_H$ (with $R_H$ the hydrodynamic radius of the knot and $\eta_0$ the
viscosity of the solvent) is smaller, resulting in higher mobilities. A more
refined calculation of the friction coefficient $\gamma$ relies on
Kirkwood-Riseman formula \cite{torr81}: 
\begin{equation}
\gamma=\left(\sum\limits_{i=1}^N{\zeta_i}\right)\left[1+\left(6\pi\eta_0
\sum\limits_{i=1}^N{\zeta_i}\right)^{-1}\sum\limits_{i=1}^N\sum%
\limits_{j=1}^N {\zeta_i\zeta_jR_{ij}^{-1}}\right]^{-1}  \label{frott}
\end{equation}
where the chain is modelled by $N$ beads of radius $\sigma_i$ and friction
coefficent $\zeta_i=6\pi\eta_o\sigma_i$, and $R_{ij}$ is the distance
between beads $i$ and $j$. The term $\frac{1}{R_{ij}}$, due to hydrodynamic
interactions between beads, in the second factor of equation (\ref{frott})
explains the observed behavior: more compact molecules have smaller
distances $R_{ij}$, and thus a smaller friction coefficient. The calculation
of an average friction coefficient $\bar{\gamma}$ on an equilibrium set of
thermally agitated DNA molecules forming different types of knots has
confirmed the experimental results \cite{volo98}.

Recently, it was observed by two-dimensional agarose gel electrophoresis
that when the strength of the electric field is increased, the
electrophoretic mobility of DNA knots changes behavior (fig.\ \ref%
{Sogo_gel_part}) \cite{roca01,stas99}. The experiment was performed in two
steps: a low strength electric field of $0.6\,V cm^{-1}$ was first applied
along one direction in the gel. At this step, more complex knots show a
higher mobility, in agreement with Kirkwood-Riseman formula. The same
procedure is repeated in a second step but with a stronger electric field ($%
5\,V cm^{-1}$) applied perpendicularly to the first one. In this case, the
opposite behavior is observed: more complex knots cover smaller distances
than simple ones.

The presence of two regimes of weak and strong electric field can be
captured with a simple argument. A knot of size $\xi$ drifts over a distance
equal to its size along the direction of the applied electric field $%
\mathcal{E}$ in a time $t_\xi \simeq\xi/v =\gamma\xi/q\mathcal{E}$, where $%
v=q\mathcal{E}/\gamma$ is the drift velocity in the stationary regime and $q$
the total electrical charge carried by the DNA molecule. During the same
time, the drifting knot diffuses laterally over a distance $d \simeq \sqrt{%
2Dt_\xi} = \sqrt{2 \frac{k_B T}{\gamma} t_\xi} = \sqrt{2 \frac{k_B T}{q%
\mathcal{E}}\xi}$, where $T$ is the absolute temperature and $k_B$ is the
Boltzmann's constant. If the transverse diffusion explores distances $d$
much larger than the typical size of the knot, $\xi$, then, on average, the
knot will be able to avoid a collision with the gel and the knots will drift
as if they were in a pure solution, with just a slight renormalization of
the friction coefficient. If instead $d\ll\xi$, then whenever a knot is on a
collision course with a gel strand, it can not avoid it. As a consequence,
after impinging over the obstacle, the knot needs to crawl around it in
order to free itself. Crawling around an obstacle is much more difficult for
more complex knots than for simple ones, due to the self-avoidance
constraint. Following this argument, the two electric field regimes are
separated by a critical field $\mathcal{E}_c$ that can be obtained by
setting $d \simeq\xi$, giving $\mathcal{E}_c \simeq 2 \frac{k_B T}{q\xi}$.
In order to estimate $\mathcal{E}_c$ we use the typical values for a $%
11^{\prime}000$ base pairs DNA knot: the size of a closed DNA ring is about $%
\xi=300\ nm$\cite{private}; the total charge $q$ depends on the gel
conditions, since every nucleotide carries a $P^-$ group, hence one electron
negative charge, that can be strongly screened by charges in the solvent. As
a consequence we use $q \sim 10^{-15}-10^{-16} C$. We then obtain a critical
field $\mathcal{E}_c=0.1-1\ V cm^{-1}$ in reasonable agreement with
experiments\cite{roca01,stas99}. The expression for the critical electric
field holds also in the case when the gel is concentrated as it is the case
in many experiments. Under the condition of high gel density, the DNA is
filling the pores and the expression for the collision condition is that the
DNA lying between two gel strands will collide with one gel strand before
the DNA can drift transversally to the electric field. Instead of $\xi$, one
has to insert the gel pore size $\ell$. The lateral diffusion constant $D^*$
has to be rescaled in order to include the effect of the gel. The condition
for the critical electric field reads again: $\mathcal{E}_c\simeq 2\frac{k_BT%
}{q\ell}$.

One has also to note that the two conditions for the critical electric field 
$\mathcal{E}_c$ actually mean that the energy gained by the DNA when moving
one diameter $\xi$ or one pore size $\ell$ along the electric field is equal
to twice the thermal energy: $q\mathcal{E}_c\xi$ (or $\ell$) $\simeq 2k_BT$.
Although the above model gives a first hint of the origin of the observed
behavior, here we want to address the issue more thoroughly using lattice
Monte Carlo simulations.

DNA knots are modelled by closed self-avoiding walks (SAWs) composed of $N$
segments of length $a$ on a three-dimensional cubic lattice (the lattice
constant $a$ is comparable to the persistence length of the DNA molecules).
The gel is a two dimensional grid forming a sublattice with a mesh size $b$
(= gel parameter) and perpendicular to the applied electric field (so that
no knots can ever get impaled). The gel lattice is shifted by the quantity $(%
\frac{a}{2},\frac{a}{2},\frac{a}{2})$ compared to the knot lattice, so that
no points of the knot lie on the gel. Knots are not allowed to cross the gel
network. The coordinates of the $N$ monomers in the configuration at time $t$
are written as: 
\begin{equation}
\bar{r}(t)=(\vec{r}_1(t),\vec{r}_2(t),...,\vec{r}_N(t))
\end{equation}
with constraints $\|\vec{r}_j(t) - \vec{r}_{j+1}(t)\| = a$.

The dynamics is followed using the BFACF algorithm~\cite{BF81ACF83}. Two
types of moves are allowed: (a) the creation/destruction of a handle and (b)
the flip of a corner into the mirror position (see Fig.\ref{BFACF moves}).
The first move clearly does not preserve the knot length, which can vary by $%
\pm 2$ at every step, but introduces the knot elasticity. The BFACF
algorithm preserves knot classes, within which it is ergodic~\cite{JvRW91}.
Self-avoidance is imposed by disallowing monomers to visit any site which is
already occupied by other monomers. Furthermore, knots are not allowed to
cross gel rods, so that corner flips and handle creation/destruction are
forbidden when a rod has to be crossed.

Under an external uniform electric field $\vec{\mathcal{E}}$, the
electrostatic energy at time $t$ is given by: 
\begin{equation}
E_q(t)=-\frac{q}{N(t)}\sum\limits_{j=1}^{N}{\vec{r}_j(t) \cdot \vec{\mathcal{%
E}}}.  \label{electrostatic energy}
\end{equation}
$N(t)$ is the length of the knot at time $t$, and it is associated with an
elastic energy 
\begin{equation}
E_{el}(t)=\frac{1}{2} K \left[N(t)-N_0\right]^2  \label{elastic energy}
\end{equation}
where $K$ is the spring constant. In the simulation a value $K/k_BT=0.1$ was
used. The knot energy is then $E(t) = E_q(t)+ E_{el}(t)$.

At each timestep, we choose a point at random on the chain and propose
alternatively one of the two moves. If it satisfies the self-avoiding and
gel-avoiding constraints, it is accepted with a probability given by the
Metropolis algorithm: if the energy of the new trial configuration, $%
E_{trial}$, is lower than that of the previous configuration, $E_{old}=E(t)$%
, the move is accepted and $\bar{r}(t+1) = \bar{r}_{trial}$; otherwise, the
probability of acceptance of the trial configuration is equal to $%
\exp\{-[E_{trial}-E(t)]\}/{k_B T}$. If the move is rejected, then $\bar{r}%
(t+1) = \bar{r}(t)$.

After a knotted configuration is randomly generated, the knot type is
obtained by calculating its Alexander polynomial~\cite{alexander}. Then, we
let the system freely relax to thermodynamic equilibrium in the absence of
an external field ($\mathcal{E}=0$) until correlations from the initial
configuration have disappeared. Then the electric field is switched on, and
we let the knots migrate on the lattice. The quantities we compute are the
position of the center-of-mass and the average crossing number (ACN) of the
knot along a trajectory.

Time is measured in Monte Carlo iterations, length in lattice spacing. The
initial length $N_0$ of our polymers was set to $150$, and the mean length
of the knot depends generally on the electric field and on the gel
parameter. However, the mean length is 145 (146) for C=0.1 (C=0.4) and b=20
and we checked that during the simulations it fluctuates around that value.
The average length is slightly shorter that $N_0$, since the probability of
shortening the polymer is a slightly larger than the probability of
lengthening it due to the self-avoiding condition. The gel parameter was set
to $b= 5, 10, 20$ (in units of $a$), corresponding to a relatively sparse
gel with big pores. For each initial knot, $20 \cdot 10^6$ iterations were
performed. The center-of-mass position has been measured every $1000$ Monte
Carlo steps, and it was then averaged over the trajectories obtained by the
migration of 100-200 different initial knots (to obtain an accuracy of about 
$10\%$).

One problem with the Monte Carlo algorithm is that more complex knots have a
smaller drift velocity than less complex ones even in absence of the gel,
when time is measured in Monte Carlo steps. This is due to the fact that
already in the absence of the gel, the moves are hindered by the complexity
of the knot. In order to correct this problem, we used Kirkwood-Riseman
formula (\ref{frott}) to compute the friction coefficient $\gamma$ of every
knot: since then $v=q\mathcal{E}/\gamma$, we can find the specific time
rescaling necessary to go from Monte Carlo time ($t_{MC}$) to real time ($t_R
$): $t_R = (\gamma v_{MC}/q\mathcal{E}) t_{MC}$, where $v_{MC}$ is the
velocity measured using $t_{MC}$. Once we find the time conversion (one for
every knot class) in the absence of the gel, then we apply it throughout our
simulations in the gel.

Let us begin with the study of the dependence of the velocity with respect
to the adimensional constant $\mathcal{C}=q\mathcal{E}a/k_B T$ for knot type 
$3_1$ (if we assume that length fluctuations are very small, energy
variations for the Metropolis algorithm can be expressed as integer
multiples of $k_B T \mathcal{C}$). We observe two distinct behaviors for the
migration of knots (see fig.\ \ref{Figure_Vitesse_E_Field}) as a function of 
$\mathcal{C}$.

At high temperature or weak electric field, the distance of migration is
linear as a function of $\mathcal{C}$. On the other hand, above a critical
value of $\mathcal{C}_{crit}$, the average speed of the knots is decreasing
with $\mathcal{C}$, in qualitative agreement with the experimental results.
For our parameters, $\mathcal{C}_{crit}$ is located around $0.4$ for the $3_1
$ knot. Clearly, this value depends on the length of the knot and on its
type (it can also depend on the gel parameter $b$ if $b<\xi$, the typical
size of the knot). If the electric field is strong, or temperature is low,
knots tend to hang over obstacles and take a U-shape configuration. When a
knot hits a gel rod, it can easily remain trapped there, because the
probability of a backward step is very small, and it is a growing function
of the temperature $T$. Trapping in U-shape conformations introduces
plateaus in the migration distance as a function of time for individual
knots, hence reducing the average migration velocity. On the same graph
(fig. \ref{Figure_Vitesse_E_Field}), the drift velocity for a $8_1$ knot is
depicted as a function of $\mathcal{C}$. The general behavior is similar to
the $3_1$ knot, but the drift speed at low $\mathcal{C}$ is higher than for
the $3_1$ knot (as it is the case in the experiments). But the most striking
feature is that at $\mathcal{C}\approx 0.2-0.3$ the two curves cross each
other and the $3_1$ is faster than the $8_1$ for $\mathcal{C}>0.3$. This is
also the observed behavior in the experiments.

We investigate now the effect of the knot type for both weak and strong
electric fields ($\mathcal{C} =0.1$ and $\mathcal{C}=0.4$ respectively) for
eleven types of knots ($3_1$, $4_1$, $5_1$, $5_2$, $6_1$, $6_2$, $6_3$, $7_1$%
, $7_2$, $7_3$ and $8_1$) all consisting of 150 monomers. For each knot
type, we extract the average velocity from the distance of migration vs.
time curve. The velocity of migration is then plotted as a function of the
measured ACN, that is related to the knot type. This plot is done first only
for the high electric field case ($\mathcal{C}=0.4$) in fig.\ \ref{vitesse}.
We observe that there is a fairly linear relationship (except for the $3_1$
knot) between the average velocity of knots and the ACN (measure of
complexity). More complex knots migrate slower than simpler ones at strong
electric fields (although much noisier than for weak fields). These results
are in agreement with experiments. A similar plot was done for the low
electric field ($\mathcal{C}=0.1$) and the results are depicted in fig. \ref%
{SogoVSSimul} (filled squares).

The intuitive view of a knot making its way through the pores of the gel
would have as a consequence that more complex (thus more compact) knots
migrate faster than simple knots, because they are less disturbed by the
gel. This is indeed what happens in weak electric fields. In the strong
field regime our simulations are in agreement with experiments and show the
opposite behavior, indicating that the knot collides with the gel and that
the condition of self-avoidance makes the migration of compact knots around
the gel strands more difficult. Somehow, parts of the knot have to go around
other parts of itself, a process which is much longer for complex knots than
for simple knots.

The trapping of linear open polymers in U-shape conformations is actually an
artifact: indeed the slightest difference in the length of the two arms of
the U gives rise to a net electric force that allows the polymer to slide
around the obstacle. Simple local Monte Carlo moves do not allow to capture
this dynamics. Yet, adding suitable non-local moves is enough to eliminate
the artificious slowing down of the dynamics and cures the exponentially
long relaxation times of thermal activation around the obstacle \cite%
{Viovy92,Barkema}. However, the closed knot topology of DNA in our numerical
simulations does not allow to introduce the long range moves. Moreover, we
argue that exponentially long relaxation times present in our simulations
affect, in first approximation, all the knots irrespective of their
topology. In the ideal case of purely mechanical and frictionless unbinding,
one can easily check that the knot complexity introduces at most a small
logarithmic correction according to which more complex knots would anyway
unpin faster in contradiction with both experiments and our simulation.
Therefore, the simulated absolute drift velocity $v_{abs}$ is affected by a
time scale that is artificially stretched in essentially the same way
independent on the knot class. So, by looking at the ratios of the absolute
velocities, this time scale should in first approximation cancel out. In
Figure \ref{SogoVSSimul} we plot the simulated and experimental ratios $%
v_{m.n}^{abs}/v_{3.1}^{abs}$ as function of the ACN. In this Figure the open
points represent data from Sogo et al.\cite{stas99}, and the filled symbols
our simulation results. The ACN values for each knot were taken from
Vologdskii et al. \cite{volo98}. The agreement between experimental data and
simulations is remarkable. In weak electric field, the transport properties
of DNA knots are dominated by the hydrodynamics of the knots and the gel
plays a minor role. At high electric fields, the knot-gel interaction is
predominant and it is responsible for the inversion of behavior. In
particular, strong DNA-gel interactions enhance the effect of self-avoidance
within the knotted DNA and self-avoidance must be included in simulations in
order to reproduce the correct behavior. This is at variance with
simulations of gel electrophoresis of linear DNA where self-avoidance is
usually neglected because of the two following reasons: first, according to
the repton model, DNA crawls along tubes in the gel and it is in an
elongated configuration where self-intersections are negligible. Secondly,
it is often assumed that the drifting DNA can be considered to be in a
semi-dilute solution regime, where polymers obey random walk statistics.
Instead, in our case the conservation of the knot topology during the
dynamics imposes the inclusion of self-avoidance.

We presented some results of a Monte Carlo simulation of DNA knots in a gel
in three dimensions. In summary, our model allows to explain the high
electric field behavior observed in experimental DNA knots gel
electrophoresis. The variation of the gel parameter does not change
qualitatively the results. However, in a denser gel, the knots get stuck at
shorter distances. In a more realistic modelization of the gel, the gel
fibers should be allowed to break and let the knots migrate further. Varying
the length of the knots would probably also bring some further insights into
the mechanisms of the migration in a gel.

We thank X. Zotos, A. Baldereschi, A. Stasiak, J. Roca, J. Schvartzman, and
G. D'Anna for help and for fruitful discussions. This work was partially
supported by the Swiss National Science Foundation (Grant Nr. 21-50805.97)

\eject \noindent Figure Captions

\begin{itemize}
\item {Figure 1: } Analysis by two-dimensional agarose gel electrophoresis
of pH5.8 DNA (8749 bp) with the diagrammatic interpretation of the
autoradiogram and the standard agarose electrophoresis (left panel).
Reproduced from \cite{stas99} with permission.

\item {Figure 2:} Elementary moves of the BFACF algorithm. a) creation of a
handle (the opposite move destroys a handle) and b) corner flip.

\item {Figure 3:} Drift velocity in arbitrary units for the $3_1$ (open
circles) and $8_1$ (filled circles) knots as function of $\mathcal{C}=q%
\mathcal{E}a/k_BT$. The lines are only guides for the eyes.

\item {Figure 4:} Linear relation between the electrophoretic drift velocity
of the centre-of-mass as a function of the average crossing number
(determined during the simulation) for different knots for low and high
electric field (filled circles for $\mathcal{C}=0.1$, open circles for $%
\mathcal{C}=0.4$).

\item {Figure 5:} Distance of migration in gel for the experimental data or
drift velocity for the simulated data vs. ACN (from \cite{volo98}) for knots
from $3_1$ to $8_1$. Open symbols are data from Sogo et al. \cite{stas99},
filled symbols are the simulated values of the drift velocity. Squares are
for low electric field ($\mathcal{E}=0.6\ V/cm$ and $\mathcal{C}=0.1$,
respectively), while triangles are for high electric field ($\mathcal{E}=5\
V/cm$ and $\mathcal{C}=0.4$). The values were all normalized to their
respective value for the knot $3_1$.
\end{itemize}

\eject

\begin{figure}[htbp]
\begin{center}
\includegraphics[width=5in]{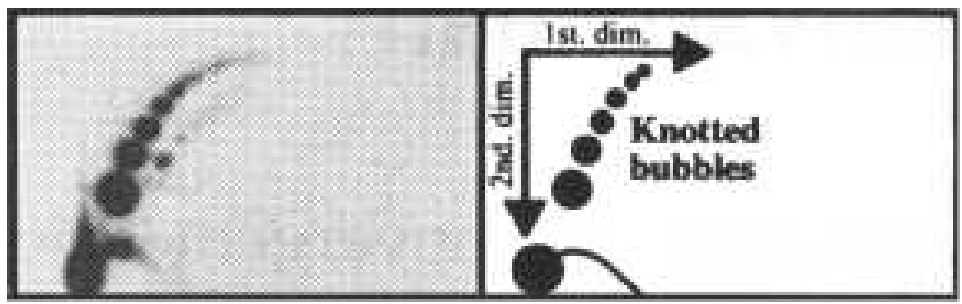}
\end{center}
\caption{Analysis by two-dimensional agarose gel electrophoresis of pH5.8
DNA (8749 bp) with the diagrammatic interpretation of the autoradiogram and
the standard agarose electrophoresis (left panel). Reproduced from 
\protect\cite{stas99} with permission.}
\label{Sogo_gel_part}
\end{figure}

\eject

\begin{figure}[htbp]
\begin{center}
\includegraphics[width=2in]{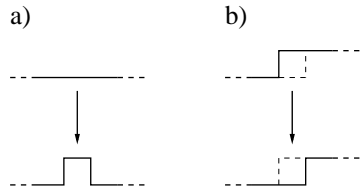}
\end{center}
\caption{{Elementary moves of the BFACF algorithm. a) creation of a handle
(the opposite move destroys a handle) and b) corner flip.}}
\label{BFACF moves}
\end{figure}

\eject

\begin{figure}[htbp]
\begin{center}
\includegraphics[width=6in]{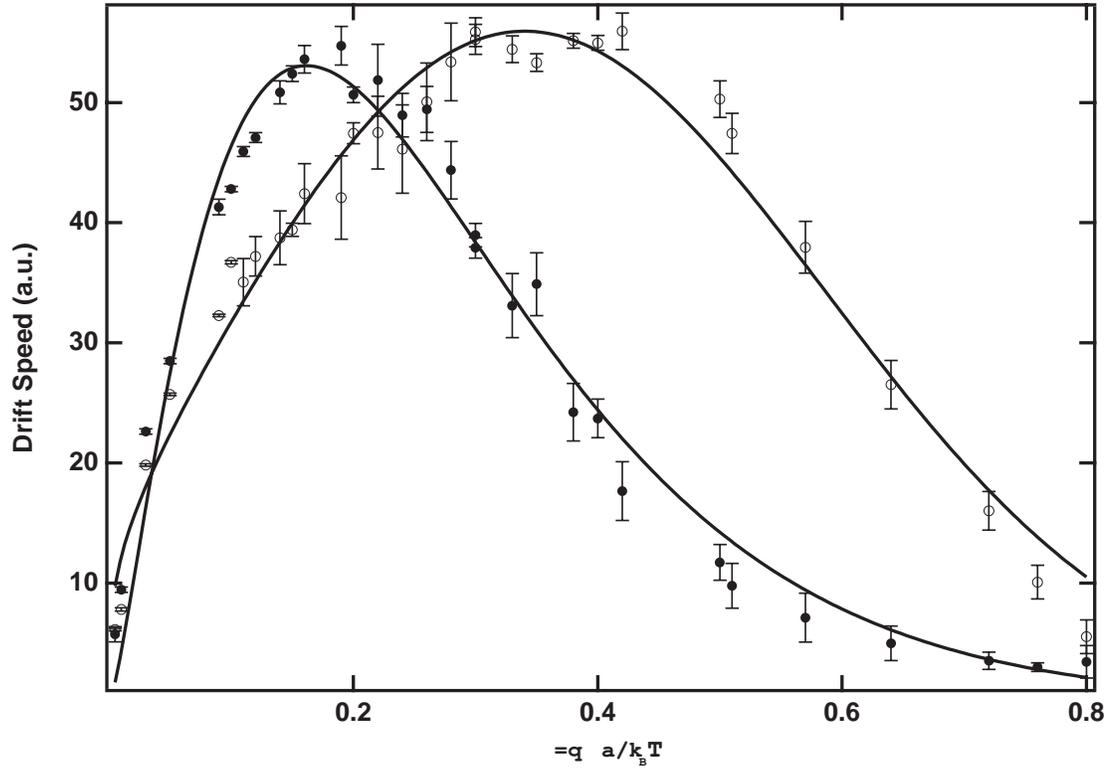}
\end{center}
\caption{{Drift velocity in arbitrary units for the $3_1$ (open circles) and 
$8_1$ (filled circles) knots as function of $\mathcal{C}=q\mathcal{E}a/k_BT$%
. The lines are only guides for the eyes.}}
\label{Figure_Vitesse_E_Field}
\end{figure}

\eject

\begin{figure}[tbp]
\includegraphics[width=6in]{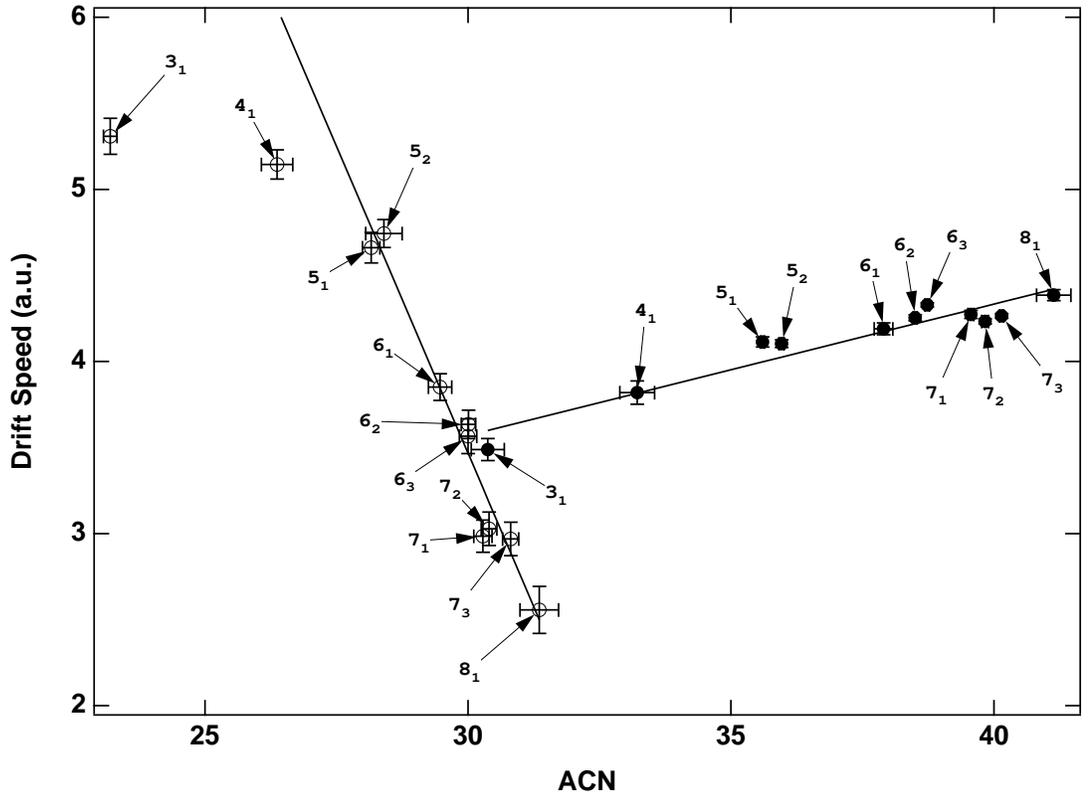}
\caption{Linear relation between the electrophoretic drift velocity of the
centre-of-mass as a function of the average crossing number (determined
during the simulation) for different knots for low and high electric field
(filled circles for $\mathcal{C}=0.1$, open circles for $\mathcal{C}=0.4$).}
\label{vitesse}
\end{figure}

\eject

\begin{figure}[tbp]
\includegraphics[width=6in]{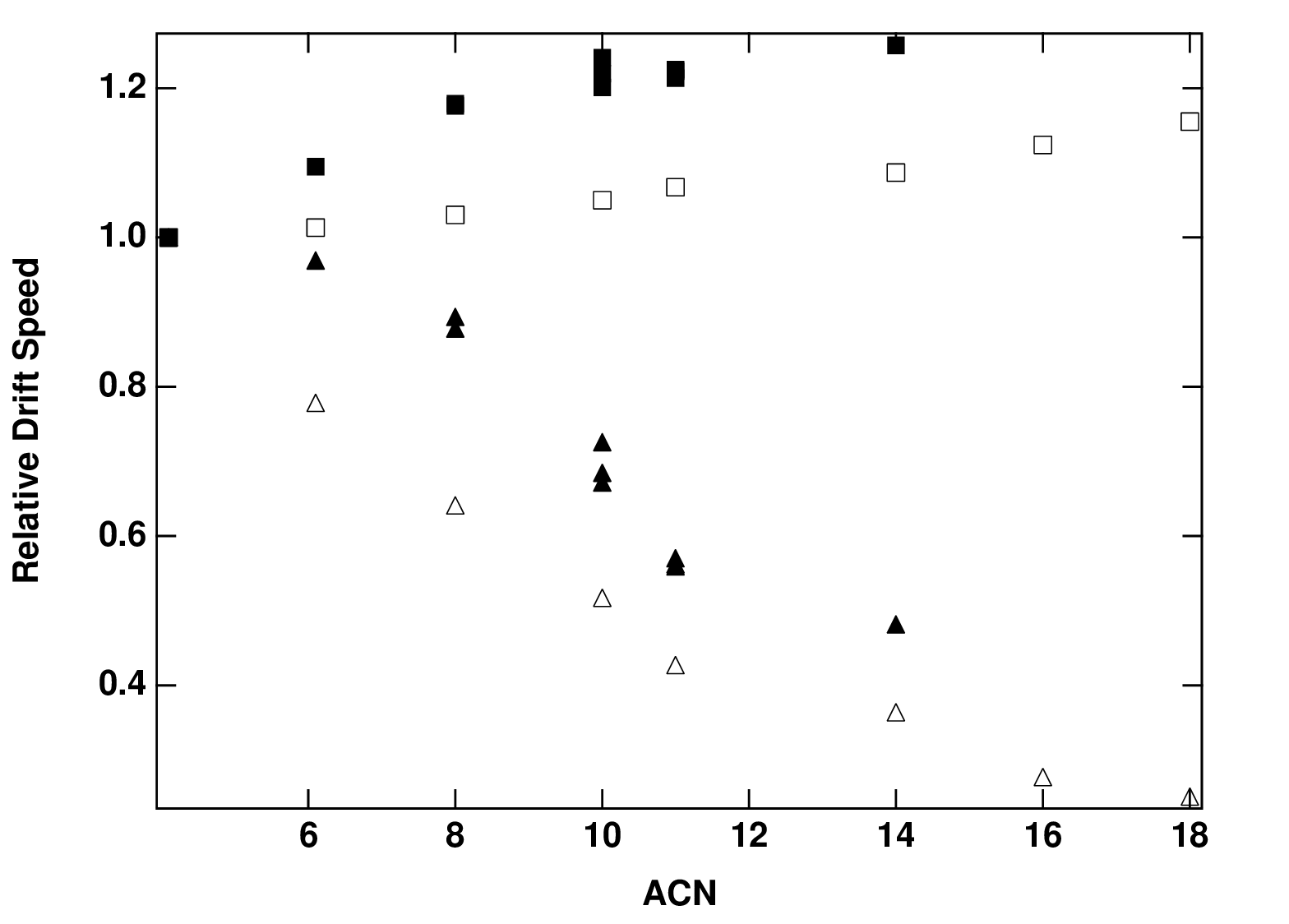}
\caption{Distance of migration in gel for the experimental data or drift
velocity for the simulated data vs. ACN (from \protect\cite{volo98}) for
knots from $3_1$ to $8_1$. Open symbols are data from Sogo et al. 
\protect\cite{stas99}, filled symbols are the simulated values of the drift
velocity. Squares are for low electric field ($\mathcal{E}=0.6\ V/cm$ and $%
\mathcal{C}=0.1$, respectively), while triangles are for high electric field
($\mathcal{E}=5\ V/cm$ and $\mathcal{C}=0.4$). The values were all
normalized to their respective value for the knot $3_1$.}
\label{SogoVSSimul}
\end{figure}


\begin{thebibliography}{*}
\bibitem{private} This value is estimated from Atomic Force Images of
complex knots. The knots were kindly provided by Dr. J. Roca and the AFM
images were taken by Dr. F. Valle.
\end{thebibliography}
\end{document}